\documentclass[nofootinbib,prd,twocolumn,showpacs,showkeys,preprintnumbers]{revtex4-1}
\usepackage{hyperref,amssymb,amsmath,mathrsfs,bm,graphicx}
\begin{document}
\title { Definition of complexity for  dynamical spherically symmetric dissipative  self--gravitating fluid distributions}
\author{L. Herrera}
\email{lherrera@usal.es}
\affiliation{Instituto Universitario de F\'isica
Fundamental y Matem\'aticas, Universidad de Salamanca, Salamanca 37007, Spain}
\author{A. Di Prisco}
\email{alicia.diprisco@ciens.ucv.ve}
\affiliation{Escuela de F\'\i sica, Facultad de Ciencias, Universidad Central de Venezuela, Caracas 1050, Venezuela}
\author{J. Ospino}
\email{j.ospino@usal.es}
\affiliation{Departamento de Matem\'atica Aplicada and Instituto Universitario de F\'isica
Fundamental y  Matem\'aticas, Universidad de Salamanca, Salamanca 37007, Spain}
\date{\today}
\begin{abstract}
The recently proposed  definition of complexity  for static   and spherically symmetric self--gravitating systems \cite{c1},  is extended to the fully dynamic situation. In this latter case we have to consider not only the complexity factor of the structure of the fluid distribution, but also the condition of minimal complexity of the pattern of evolution. As we shall see these two issues are deeply intertwined.  For the complexity factor of the structure we choose the same as for the static case, whereas for the simplest pattern of evolution we assume the homologous condition. The dissipative and non-dissipative cases are considered separately. In the latter case the fluid distribution, satisfying the vanishing complexity factor condition and evolving homologously, corresponds to a homogeneous (in the energy  density), geodesic and shear--free,  isotropic (in the pressure) fluid. In the dissipative case the fluid is still geodesic, but shearing, and there exists (in principle) a large class of solutions. Finally we discuss about the stability of the vanishing complexity condition.

 \end{abstract}
\date{\today}
\maketitle

\section{Introduction}
In a recent paper   a new definition of complexity for spherically symmetric  static self--gravitating fluids, in the context of general relativity, has been introduced \cite{c1} (for an extension of this concept to other theories of gravitation see\cite{ab}).

The new defined variable is sharply different  from a previous definition given in  \cite{5, 6, 7, 8, 9, 10}  which  was based on the work developed by L\'opez--Ruiz and collaborators \cite{3, 4}.

The new concept of complexity, for static spherically symmetric relativistic fluid distributions, stems from the basic assumption that one of the less complex systems corresponds to an homogeneous (in the energy density) fluid distribution with isotropic pressure. So we assign a zero value of  the complexity factor for such a distribution. Then, as an obvious candidate to measure the degree of complexity, emerges a quantity that  appears   in the orthogonal splitting of the Riemann tensor and that was denoted by $Y_{TF}$  and called the complexity factor.  

The main  reason behind such a proposal resides in the fact that the scalar function $Y_{TF}$  contains contributions from the energy density inhomogeneity and the local pressure anisotropy, combined in a very specific way, which vanishes for the homogeneous and locally isotropic fluid distribution.  Also as shown in Appendix A,  in the case of a charged fluid,  this scalar also encompasses the effect of the electric charge.

It is worth mentioning that the complexity factor so defined,  not only vanishes for the simple configuration mentioned above, but also may vanish when the two terms appearing in its definition, and  containing density inhomogeneity and anisotropic pressure, cancel each other. Thus as in \cite{3}, vanishing complexity  may correspond to very different systems.

Once the complexity factor is defined for static fluid distributions, the obvious question arises: how to define complexity for dynamical self--gravitating systems? It is the purpose of this work to answer  such  question.

When dealing with time dependent  systems we face two different problems; on the one hand we have to generalize the concept of complexity  of the structure of the fluid distribution to time dependent dissipative fluids, and on the other hand we also have  to evaluate the complexity of the patterns of evolution and propose what we consider is the simplest of them. 

As we shall see here, the complexity factor for the structure of the fluid distribution is the same scalar function  $Y_{TF}$, which now includes the dissipative variables. As for the simplest pattern of evolution we shall start by considering two possible modes of evolution: homogeneous expansion and homologous evolution. For reasons that will be explained below
we shall assume the homologous condition as the one characterizing the simplest mode of evolution.

The imposition of the vanishing complexity factor, and the homologous evolution leads to a geodesic fluid. If, we further assume the fluid to be non--dissipative (but time dependent), then it will  also be shear--free, endowed with a homogeneous energy--density and isotropic pressure. Also, in this case the homogeneous expansion  and the homologous condition imply each other, and there is a unique solution fulfilling the minimal complexity criteria. In the most general (dissipative) case the fluid is shearing and there exist a large family of solutions.

Finally we discuss about the stability of the vanishing complexity condition, and find the physical factors that could produce deviations from such a condition.

The paper is organized as follows. In the next section we present the  notation, the general equations and the variables required for our discussion. In section III the complexity factor is defined, whereas in section IV we discuss about the simplest pattern of evolution (homologous and homogeneous expansion conditions).  The main consequences derived from the vanishing complexity factor condition  and the homologous evolution are discussed in sections V and  VI. The stability of the vanishing complexity factor is  tackled in section VII, and all the obtained results and open issues are summarized in the last section. Finally some useful formulae are given in the appendixes.

\section{The general setup of the problem: notation, variables and equations}
We consider a spherically symmetric distribution  of collapsing
fluid, which may be  bounded by a spherical surface $\Sigma$, or not. The fluid is
assumed to be locally anisotropic (principal stresses unequal) and undergoing dissipation in the
form of heat flow (diffusion approximation). 

Choosing comoving coordinates, the general
interior metric can be written
\begin{equation}
ds^2=-A^2dt^2+B^2dr^2+R^2(d\theta^2+\sin^2\theta d\phi^2),
\label{1}
\end{equation}
where $A$, $B$ and $R$ are functions of $t$ and $r$ and are assumed
positive. We number the coordinates $x^0=t$, $x^1=r$, $x^2=\theta$
and $x^3=\phi$. Observe that $A$ and $B$ are dimensionless, whereas $R$ has the same dimension as $r$.

The matter energy-momentum $T_{\alpha\beta}$ of the fluid distribution
has the form
\begin{eqnarray}
T_{\alpha\beta}&=&(\mu +
P_{\perp})V_{\alpha}V_{\beta}+P_{\perp}g_{\alpha\beta}+(P_r-P_{\perp})\chi_{
\alpha}\chi_{\beta}\nonumber \\&+&q_{\alpha}V_{\beta}+V_{\alpha}q_{\beta}
, \label{3}
\end{eqnarray}
where $\mu$ is the energy density, $P_r$ the radial pressure,
$P_{\perp}$ the tangential pressure, $q^{\alpha}$ the heat flux, $V^{\alpha}$ the four velocity of the fluid,
and $\chi^{\alpha}$ a unit four vector along the radial direction. These quantities
satisfy
\begin{eqnarray}
V^{\alpha}V_{\alpha}=-1, \;\; V^{\alpha}q_{\alpha}=0, \;\; \chi^{\alpha}\chi_{\alpha}=1,\;\;
\chi^{\alpha}V_{\alpha}=0.
\end{eqnarray}

It will be convenient to express the  energy momentum tensor  (\ref{3})  in the equivalent (canonical) form
\begin{equation}
T_{\alpha \beta} = {\mu} V_\alpha V_\beta + P h_{\alpha \beta} + \Pi_{\alpha \beta} +
q \left(V_\alpha \chi_\beta + \chi_\alpha V_\beta\right) \label{Tab}
\end{equation}
with
$$ P=\frac{P_{r}+2P_{\bot}}{3}, \qquad h_{\alpha \beta}=g_{\alpha \beta}+V_\alpha V_\beta,$$

$$\Pi_{\alpha \beta}=\Pi\left(\chi_\alpha \chi_\beta - \frac{1}{3} h_{\alpha \beta}\right), \qquad \Pi=P_{r}-P_{\bot}.$$

Since we are considering comoving observers, we have
\begin{eqnarray}
V^{\alpha}&=&A^{-1}\delta_0^{\alpha}, \;\;
q^{\alpha}=qB^{-1}\delta^{\alpha}_1, \;\;
\chi^{\alpha}=B^{-1}\delta^{\alpha}_1, \label{5}
\end{eqnarray}
where $q$ is a function of $t$ and $r$.

It is worth noticing that we do not explicitly add bulk or shear viscosity to the system because they
can be trivially absorbed into the radial and tangential pressures, $P_r$ and
$P_{\perp}$, of the collapsing fluid (in $\Pi$). Also we do not explicitly  introduce  dissipation in the free streaming approximation since it can be absorbed in $\mu, P_r$ and $q$. Finally, let us mention  that the complexity factor can be  extended to the charged case, as shown in  Appendix A (a detailed analysis of the charged static case is given in \cite{sb}).

The Einstein equations for (\ref{1}) and (\ref{Tab}), are explicitly written  in Appendix B.

The acceleration $a_{\alpha}$ and the expansion $\Theta$ of the fluid are
given by
\begin{equation}
a_{\alpha}=V_{\alpha ;\beta}V^{\beta}, \;\;
\Theta={V^{\alpha}}_{;\alpha}. \label{4b}
\end{equation}
and its  shear $\sigma_{\alpha\beta}$ by
\begin{equation}
\sigma_{\alpha\beta}=V_{(\alpha
;\beta)}+a_{(\alpha}V_{\beta)}-\frac{1}{3}\Theta h_{\alpha\beta}.
\label{4a}
\end{equation}

From  (\ref{4b}) with (\ref{5}) we have for the  acceleration and its scalar $a$,
\begin{equation}
a_1=\frac{A^{\prime}}{A}, \;\; a=\sqrt{a^{\alpha}a_{\alpha}}=\frac{A^{\prime}}{AB}, \label{5c}
\end{equation}
and for the expansion
\begin{equation}
\Theta=\frac{1}{A}\left(\frac{\dot{B}}{B}+2\frac{\dot{R}}{R}\right),
\label{5c1}
\end{equation}
where the  prime stands for $r$
differentiation and the dot stands for differentiation with respect to $t$.
With (\ref{5}) we obtain
for the shear (\ref{4a}) its non zero components
\begin{equation}
\sigma_{11}=\frac{2}{3}B^2\sigma, \;\;
\sigma_{22}=\frac{\sigma_{33}}{\sin^2\theta}=-\frac{1}{3}R^2\sigma,
 \label{5a}
\end{equation}
and its scalar
\begin{equation}
\sigma^{\alpha\beta}\sigma_{\alpha\beta}=\frac{2}{3}\sigma^2,
\label{5b}
\end{equation}
where
\begin{equation}
\sigma=\frac{1}{A}\left(\frac{\dot{B}}{B}-\frac{\dot{R}}{R}\right).\label{5b1}
\end{equation}

Next, the mass function $m(t,r)$ introduced by Misner and Sharp
\cite{Misner}  reads
\begin{equation}
m=\frac{R^3}{2}{R_{23}}^{23}
=\frac{R}{2}\left[\left(\frac{\dot R}{A}\right)^2-\left(\frac{R^{\prime}}{B}\right)^2+1\right].
 \label{17masa}
\end{equation}

Introducing the proper time derivative $D_T$
given by
\begin{equation}
D_T=\frac{1}{A}\frac{\partial}{\partial t}, \label{16}
\end{equation}
we can define the velocity $U$ of the collapsing
fluid  as the variation of the areal radius with respect to proper time, i.e.
\begin{equation}
U=D_TR<0 \;\; \mbox{(negative in the case of collapse)}, \label{19}
\end{equation}
where $R$ defines the areal radius of a spherical surface inside the fluid distribution (as
measured from its area).

Then (\ref{17masa}) can be rewritten as
\begin{equation}
E \equiv \frac{R^{\prime}}{B}=\left(1+U^2-\frac{2m}{R}\right)^{1/2}.
\label{20x}
\end{equation}
Using (\ref{20x}) we can express (\ref{17a}) as
\begin{equation}
4\pi q=E\left[\frac{1}{3}D_R(\Theta-\sigma)
-\frac{\sigma}{R}\right],\label{21a}
\end{equation}
where   $D_R$ denotes the proper radial derivative,
\begin{equation}
D_R=\frac{1}{R^{\prime}}\frac{\partial}{\partial r}.\label{23a}
\end{equation}
Using (\ref{12})-(\ref{14}) with (\ref{16}) and (\ref{23a}) we obtain from
(\ref{17masa})
\begin{eqnarray}
D_Tm=-4\pi\left(P_rU+ qE\right)R^2,
\label{22Dt}
\end{eqnarray}
and
\begin{eqnarray}
D_Rm=4\pi\left(\mu+q\frac{U}{E}\right)R^2,
\label{27Dr}
\end{eqnarray}
which implies
\begin{equation}
m=4\pi\int^{r}_{0}\left( \mu +q\frac{U}{E}\right)R^2R^\prime dr, \label{27intcopy}
\end{equation}
satisfying the regular condition  $m(t,0)=0$.

Integrating (\ref{27intcopy}) we find
\begin{equation}
\frac{3m}{R^3} = 4\pi {\mu} - \frac{4\pi}{R^3} \int^r_0{R^3\left(D_R{ \mu}-3 q \frac{U}{RE}\right) R^\prime dr}.
\label{3m/R3}
\end{equation}
\\

\subsection{The structure scalars}
 As we shall see below, the complexity factor, as for the static case, will be represented by a scalar function belonging to a set of variables denoted as structure scalars, and which appear in orthogonal splitting of the Riemann tensor. Such scalar functions were defined in \cite{20}. Here we shall briefly review the process of their obtention.
 
Let us first recall that in  the spherically symmetric case the Weyl tensor  ($C^{\rho}_{\alpha
\beta
\mu}$) is   defined by its ``electric'' part 
 $E_{\gamma \nu }$, since its  ``magnetic'' part 
vanishes, 
  \begin{equation}
E_{\alpha \beta} = C_{\alpha \mu \beta \nu} V^\mu V^\nu,
\label{elec}
\end{equation}
whose non trivial components are
\begin{eqnarray}
E_{11}&=&\frac{2}{3}B^2 {\cal E},\nonumber \\
E_{22}&=&-\frac{1}{3} R^2 {\cal E}, \nonumber \\
E_{33}&=& E_{22} \sin^2{\theta},
\label{ecomp}
\end{eqnarray}
where
\begin{widetext}
\begin{eqnarray}
{\cal E}= \frac{1}{2 A^2}\left[\frac{\ddot R}{R} - \frac{\ddot B}{B} - \left(\frac{\dot R}{R} - \frac{\dot B}{B}\right)\left(\frac{\dot A}{A} + \frac{\dot R}{R}\right)\right]+ \frac{1}{2 B^2} \left[\frac{A^{\prime\prime}}{A} - \frac{R^{\prime\prime}}{R} + \left(\frac{B^{\prime}}{B} + \frac{R^{\prime}}{R}\right)\left(\frac{R^{\prime}}{R}-\frac{A^{\prime}}{A}\right)\right] - \frac{1}{2 R^2}.
\label{E}
\end{eqnarray}
\end{widetext}
 Observe that  the electric part of the 
Weyl tensor, may be written as:
\begin{equation}
E_{\alpha \beta}={\cal E} (\chi_\alpha \chi_\beta-\frac{1}{3}h_{\alpha \beta}).
\label{52}
\end{equation}
As we mention in the Introduction, the scalar function $Y_{TF}$ appears in a natural way in the orthogonal splitting of the Riemann tensor (see \cite{20} for details).

Indeed, 
let us define tensors $Y_{\alpha \beta}$ and 
$X_{\alpha \beta}$ by:
\begin{equation}
Y_{\alpha \beta}=R_{\alpha \gamma \beta \delta}V^\gamma V^\delta,
\label{electric}
\end{equation}
\begin{equation}
X_{\alpha \beta}=^*R^{*}_{\alpha \gamma \beta \delta}V^\gamma
V^\delta=\frac{1}{2}\eta_{\alpha\gamma}^{\quad \epsilon
\rho}R^{*}_{\epsilon \rho\beta\delta}V^\gamma V^\delta,
\label{magnetic}
\end{equation}
where $R^*_{\alpha \beta \gamma \delta}=\frac{1}{2}\eta
_{\epsilon \rho \gamma \delta} R_{\alpha \beta}^{\quad \epsilon
\rho}$.

Tensors $Y_{\alpha \beta}$ and  $X_{\alpha \beta}$ may be expressed  in terms of four scalar functions $Y_T, Y_{TF}, X_T, X_{TF}$ (structure scalars) as: 
\begin{eqnarray}
Y_{\alpha\beta}=\frac{1}{3}Y_T h_{\alpha
\beta}+Y_{TF}(\chi_{\alpha} \chi_{\beta}-\frac{1}{3}h_{\alpha
\beta}),\label{electric'}
\\
X_{\alpha\beta}=\frac{1}{3}X_T h_{\alpha
\beta}+X_{TF}(\chi_{\alpha} \chi_{\beta}-\frac{1}{3}h_{\alpha
\beta}).\label{magnetic'}
\end{eqnarray}

Then after lengthy but simple calculations, using field equations [see (23), (24) in \cite{21}], and (\ref{E}) we obtain
\begin{eqnarray}
Y_T=4\pi(\mu+3 P_r-2\Pi) , \qquad
Y_{TF}={\cal E}-4\pi \Pi ,\label{EY}
\\
X_T=8\pi \mu , \qquad
X_{TF}=-{\cal E}-4\pi \Pi.\label{EX}
\end{eqnarray}
  
Next, using  (\ref{12}), (\ref{14}), (\ref{15}) with (\ref{17masa}) and (\ref{E}) we obtain
\begin{equation}
\frac{3m}{R^3}=4\pi \left({\mu}-\Pi \right) - \cal{E},
\label{mE}
\end{equation}
which combined with (\ref{3m/R3})  and (\ref{EY}) produces

\begin{equation}
Y_{TF}= -8\pi\Pi +\frac{4\pi}{R^3}\int^r_0{R^3\left(D_R {\mu}-3{q}\frac{U}{RE}\right)R^\prime dr}.
\label{Y}
\end{equation}
It is worth noticing that due to a different signature, the sign of $Y_{TF}$ in the above equation differs from the sign of the $Y_{TF}$ used in \cite{c1} for the static case.

Thus the scalar $Y_{TF}$ may be expressed through the Weyl tensor and the anisotropy of pressure  or in terms of the anisotropy of pressure, the density inhomogeneity and  the dissipative variables. 

From the above it also follows that
\begin{equation}
X_{TF}=-\frac{4\pi}{R^3}\int^r_0{R^3\left(D_R {\mu}-3{q}\frac{U}{RE}\right)R^\prime dr}.
\label{X}
\end{equation}

Finally, a differential equation for the Weyl tensor and the energy density inhomogeneity can be written as 
\begin{equation}
\left(X_{TF}+4\pi \mu \right)^\prime=  - X_{TF} \frac{3R^\prime}{R} +4 \pi q  B (\Theta - \sigma).
\label{wpxeff}
\end{equation}
 [see (37) in \cite{H5}].

From the above equation, it follows at once that in the non--dissipative case
 \begin{equation}
X_{TF}=0\Leftrightarrow  \mu^\prime=0,
\label{ief1}
\end{equation}
whereas in the general dissipative case
 \begin{equation}
X_{TF}=0\Leftrightarrow  \mu^\prime= qB(\Theta-\sigma)= qB\frac{3\dot R}{R}.
\label{ief2}
\end{equation}
\
\subsection{The exterior spacetime and junction conditions}
In the case that we consider bounded fluid distributions, then we still have to satisfy the junction (Darmois) conditions. Thus, outside $\Sigma$ we assume we have the Vaidya
spacetime (i.e.\ we assume all outgoing radiation is massless),
described by
\begin{equation}
ds^2=-\left[1-\frac{2M(v)}{r}\right]dv^2-2drdv+r^2(d\theta^2
+\sin^2\theta
d\phi^2) \label{1int},
\end{equation}
where $M(v)$  denotes the total mass,
and  $v$ is the retarded time.

Thus the matching of the full nonadiabatic sphere   to
the Vaidya spacetime, on the surface $r=r_{\Sigma}=$ constant, requires
 \begin{equation}
m(t,r)\stackrel{\Sigma}{=}M(v), \label{junction1}
\end{equation}
and
\begin{widetext}
\begin{eqnarray}
2\left(\frac{{\dot R}^{\prime}}{R}-\frac{\dot B}{B}\frac{R^{\prime}}{R}-\frac{\dot R}{R}\frac{A^{\prime}}{A}\right)
\stackrel{\Sigma}{=}-\frac{B}{A}\left[2\frac{\ddot R}{R}
-\left(2\frac{\dot A}{A}
-\frac{\dot R}{R}\right)\frac{\dot R}{R}\right]+\frac{A}{B}\left[\left(2\frac{A^{\prime}}{A}
+\frac{R^{\prime}}{R}\right)\frac{R^{\prime}}{R}-\left(\frac{B}{R}\right)^2\right],
\label{j2}
\end{eqnarray}
\end{widetext}
where $\stackrel{\Sigma}{=}$ means that both sides of the equation
are evaluated on $\Sigma$.

Comparing (\ref{j2}) with  (\ref{13}) and (\ref{14}) one obtains
\begin{equation}
q\stackrel{\Sigma}{=}P_r.\label{j3}
\end{equation}
Thus   the matching of
(\ref{1})  and (\ref{1int}) on $\Sigma$ implies (\ref{junction1}) and  (\ref{j3}).
\section{The complexity factor}

In the present case the definition of a quantity measuring the complexity of the system poses two  additional problems with respect to the static case considered  in \cite{c1}. Indeed, on the one hand, we have to deal  with the complexity of the structure of the object, which in the static case depends  only on the energy density inhomogeneity and the pressure anisotropy, but  in the case under consideration should also involve  dissipative variables.  On the other hand, we have to consider the complexity of the pattern of evolution of the system.

 For a static fluid distribution it was assumed in \cite{c1} that the simplest system is represented by a homogeneous (in the energy density), locally  isotropic fluid (principal stresses equal). So  a zero value of  the complexity factor was assumed for such a distribution. Furthermore it was shown that   Tolman mass, which may be interpreted as the ``active''  gravitational mass, may be expressed, for an arbitrary distribution,  through its value for the zero complexity case plus two terms depending on the energy density inhomogeneity and pressure anisotropy, respectively.  These latter terms in its turn may be expressed through a single scalar function which  turned out to  be the scalar function $Y_{TF}$, and accordingly was identified as  the complexity factor. 
 
 We shall consider here that $Y_{TF}$ still measures the complexity of the system, in what corresponds to the structure of the object, and we shall adopt an assumption about the simplest possible pattern of evolution. Specifically, we shall assume that the simplest   evolution pattern (one of them at least) is described by the homologous evolution.

\section{The homologous evolution and the homogeneous expansion condition}
Once the complexity factor for the structure of the fluid distribution has been established, it remains to elucidate what is the simplest pattern of evolution.
Based on purely intuitive thoughts we can identify two patterns of evolution that might be considered as the simplest ones, and these are the homologous evolution and the homogeneous expansion ($\Theta^\prime=0$). As we shall see below, both modes of evolution imply each other in the non--dissipative case. In the most general, dissipative, case,  the arguments presented in the next section lead us to choose  the homologous evolution as the simplest mode.
\subsection{The homologous evolution}
First of all observe that  we can write (\ref{21a}) as
\begin{equation}
D_R\left(\frac{U}{R}\right)=\frac{4 \pi}{E} q+\frac{\sigma}{R},
\label{vel24}
\end{equation}
which after integration  becomes
\begin{equation}
U=a(t) R+R\int^r_0\left(\frac{4\pi}{E} q+\frac{\sigma}{R}\right)R^{\prime}dr,
\label{25}
\end{equation}
where $a$ is an integration function, or,
\begin{equation}
U=\frac{U_{\Sigma}}{R_{\Sigma}}R-R\int^{r_{\Sigma}}_r\left(\frac{4\pi}{E} q+\frac{\sigma}{ R}\right)R^{\prime}dr.
\label{26}
\end{equation}
If the integral in the above equations vanishes  we have from (\ref{25}) or (\ref{26}) that $U\sim R$, which is characteristic of the homologous evolution \cite{7'}. This may occur if the fluid is shear--free and non dissipative, or if the two terms in the integral cancel each other.

This implies that for two concentric shells of areal radii, say  $R_I$ and $R_{II}$, we have in this case
\begin{equation}
\frac{R_I}{R_{II}}=\mbox{constant}.
\label{vel22}
\end{equation}
The equation above strongly suggests that  the pattern of evolution associated with the homologous condition is the simplest (at least one of them) we could find during the evolution of the fluid distribution.

Thus, if the evolution is homologous, then 
\begin{equation}
 U=a(t)R,\qquad a(t)\equiv \frac{U_\Sigma}{R_\Sigma},
 \label{hom1}
 \end{equation}
from which it follows that $R$ is a separable function, i.e. we can write
\begin{equation}
 R=R_1(t) R_2(r).
 \label{hom2}
 \end{equation}

The second term on the right of (\ref{26}) describes how the shear and dissipation deviate the evolution from the homologous regime.

To summarize, the homologous condition  (\ref{vel22}), implies (\ref{hom2}), and 
\begin{equation}
\frac{4\pi}{R^\prime}B  q+\frac{\sigma}{ R}=0,
\label{ch1}
\end{equation}
where (\ref{20x}) has been used.
\subsection{The homogeneous expansion}

Another  pattern of evolution that could be identified as ``simple'' is described by a homogeneous expansion, which  because of (\ref{21a}) implies
\begin{equation}
4 \pi q=-\frac{R^\prime}{B}\left[\frac{1}{3}D_R(\sigma)
+\frac{\sigma}{R}\right].\label{ch2}
\end{equation}

From the above it follows that if we impose both conditions [i.e. (\ref{ch1}) and (\ref{ch2})] we get  $D_R(\sigma)=0$, which implies because of the regularity conditions in the neighborhood of the center, that $\sigma=0$, i.e. that we have no dissipation.

From (\ref{ch2}) it follows at once that  if the fluid is shearfree and the expansion scalar is homogeneous, then the fluid is necessarily non--dissipative. In this case as it follows from (\ref{26}), the fluid is also homologous.

\section{Some kinematical considerations.}

As we have seen,  the homologous condition (\ref{ch1}) reads
\begin{equation}
4\pi B q=-\frac{\sigma R^\prime}{R}.
\label{con4}
\end{equation}
Feeding back this last expression into (\ref{17a}), we obtain 
\begin{equation}
(\Theta-\sigma)^\prime=0,
\label{con5b}
\end{equation}
whereas, using (\ref{5c1}) and (\ref{5b1}) we get
\begin{equation}
\left(\Theta-\sigma\right)^\prime =\left(\frac{3}{A}\frac{\dot R}{R}\right)^\prime=0.
\label{con6b} 
\end{equation}
Then using (\ref{hom2}) it follows at once that 
\begin{equation}
A^\prime=0,
\label{con7}
\end{equation}
implying that  the fluid is geodesic, as it follows from (\ref{5c}). Also,  by reparametrizing the coordinate $r$, we may put, without loss of generality, $A=1$.

On the other hand, the inverse is also true, i.e. the geodesic condition implies that the fluid is homologous.
Indeed, from the geodesic condition we have $A=1$, implying
\begin{equation}
\Theta-\sigma=3\frac{\dot R}{R}.
\label{con9b} 
\end{equation}
Evaluating this last equation close to the center where $R\sim r$ we obtain that $(\Theta-\sigma)^\prime=0$ (close to the center). 
Taking successive $r$-derivatives of (\ref{con9b}) we obtain that close to the center
\begin{equation}
\frac{\partial^n(\Theta-\sigma)}{\partial r^n}=0,
\label{con9B} 
\end{equation}
for any $n>0$. Then assuming that  $(\Theta-\sigma)^\prime$ is of class $C^\omega$, i.e. that it equals its Taylor series expansion around the center, we can analytically continue the zero value at the center to the whole configuration, recovering (\ref{con5b}), which implies that the fluid is homologous.

Thus the homologous condition and the geodesic condition imply each other.

In the non--dissipative case, the homologous condition not only implies that the fluid is geodesic, but also that it is shear--free, as it follows at once from (\ref{26}) or (\ref{ch1}). Of course, in this case (non--dissipative), the shear--free condition also implies the homologous condition.

Let us now take a look at the homogeneous expansion condition in the non--dissipative case. This condition implies because of  (\ref{ch2}), 
\begin{equation}
\frac{\sigma^\prime}{\sigma}=-\frac{3R^\prime}{R},
\label{1con}
\end{equation}
which after integration produces 
\begin{equation}
\sigma=\frac{f(t)}{R^3},
\label{2con}
\end{equation}
where $f(t)$ is an arbitrary integration function.

Since at the origin $r=0$ we have $R=0$, it follows that we must put $f=0$, implying that $\sigma=0$.

On the other hand, we see at once from (\ref{17a}) that if $\sigma=0$ then $\Theta^\prime=0$.

To summarize, we have   that in the non--dissipative case
\begin{equation}
\sigma=0  \Leftrightarrow U\sim R  \Leftrightarrow  \Theta^\prime=0.
\label{3con}
\end{equation}

Therefore in the non--dissipative case the homologous condition and the homogeneous expansion condition imply each other. Accordingly in this particular case (non dissipative) the criterion to define the simplest pattern of evolution is unique.

An  important point  to mention here is that as we have shown in \cite{26}, an initially  shear--free geodesic fluid remains shear--free and geodesic during the evolution iff $Y_{TF}=0$. Therefore, if we consider a system that starts its evolution from the rest ($\sigma=0$), it will remain shear--free if  the fluid is geodesic (or equivalently, homologous) and $Y_{TF}=0$. This is an additional argument to support our choice of $Y_{TF}$ as the complexity factor.

Let us now turn to the following question: how is the homogeneous expansion condition related to the shear (in the general dissipative case)?

If we assume that $\Theta^\prime=0$, then equation (\ref{17a}) becomes
\begin{equation}
\sigma^\prime+\frac{3\sigma R^\prime}{R}+12\pi  q B=0,
\label{con9}
\end{equation}
whose solution is
\begin{equation}
\sigma=-\frac{12\pi \int^r_0{R^3  q B d\bar r}}{R^3}.
\label{con11}
\end{equation}

The above expression is incompatible with the homologous condition, unless we assume $q=\sigma=0$, as can be seen very easily by using (\ref{con4}) in (\ref{con11}). This, of course, is consistent with our previous remarks about the impossibility of imposing simultaneously the homologous and the homogeneous expansion conditions,  in the presence of dissipation.

\section{Some dynamical  considerations}

We have seen that the homologous condition implies that the fluid is geodesic, even in the general dissipative case. Then if we impose the homologous condition,  the equation (\ref{28}) becomes
\begin{equation}
D_TU=-\frac{m}{R^2}-4\pi  P_r R
. \label{28bis}
\end{equation}

The above equation may be written in terms of $Y_{TF}$ as
\begin{equation}
\frac{3D_TU}{R}=-4\pi\left( \mu +3 P_r- 2\Pi\right)+Y_{TF},
 \label{29bis}
\end{equation}
where (\ref{EY}) has been used.
 
Using  (\ref{12}), (\ref{14}), (\ref{15}) with (\ref{17masa}) and (\ref{E}) we obtain
\begin{equation}
\frac{3m}{R^3}=4\pi \left({\mu}-\Pi \right) - \cal{E}.
\label{mE}
\end{equation}
Next from the field equations we obtain

\begin{equation}
4\pi\left(\mu +3P_r- 2\Pi\right)=-\frac{2 \ddot R}{R}-\frac{ \ddot B}{B},
 \label{31bis}
\end{equation}
and from the definition of $U$
\begin{equation}
\frac{3D_TU}{R}=\frac{3 \ddot R}{R},
 \label{32bis}
\end{equation}
feeding back the two equations above into (\ref{29bis}), it follows that
\begin{equation}
\frac{ \ddot R}{R}-\frac{ \ddot B}{B}=Y_{TF}.
 \label{33bisa}
\end{equation}
Since we are assuming the fluid to be homologous, then using (\ref{hom1}), we can write  (\ref{29bis})  as

\begin{equation}
3\left(\dot a(t)+a(t)\frac{\dot R}{R}\right)=-4\pi\left( \mu +3P_r- 2\Pi\right)+Y_{TF}.
 \label{30bis}
\end{equation}

In the case $Y_{TF}=0$, the integration of  (\ref{33bisa}) produces
\begin{equation}
B=R_1(t)\left(b_1(r)\int{\frac{dt}{R_1(t)^2}+b_2(r)}\right),
\label{int1}
\end{equation}
where $b_1(r)$ and $b_2(r)$ are two functions of integration.

It will be convenient to write the above equation as
\begin{equation}
B=R_1(t)R^\prime_2(r)\left(\tilde b_1(r)\int{\frac{dt}{R_1(t)^2}+\tilde b_2(r)}\right),
\label{int1bis}
\end{equation}
with $ b_1(r)=\tilde b_1(r)R_2^\prime$ and $ b_2(r)=\tilde b_2(r)R_2^\prime$.

Then introducing the variable 
\begin{equation}
Z=\tilde b_1(r)\int{\frac{dt}{R_1(t)^2}+\tilde b_2(r)},
\label{int13}
\end{equation}
we may write
\begin{equation}
B=ZR^\prime.
\label{int14}
\end{equation}
Let us now consider first the non--dissipative case.

\subsection{The non--dissipative case}
If we further assume  the fluid to be non--dissipative, 
then recalling that in this case the homologous condition implies the vanishing of the shear, we have because  of (\ref{5b1}) and (\ref{3con}) 
\begin{equation}
\frac{ \ddot R}{R}-\frac{ \ddot B}{B}=0\quad \Rightarrow Y_{TF}=0.
 \label{33bis}
\end{equation}

In other words, in this particular case, the homologous condition already implies the vanishing complexity factor condition.

Furthermore, since  the fluid is shear--free, we have because of (\ref{5b1}) and (\ref{int1})
\begin{equation}
b_1(r)=0 \Rightarrow B=b_2(r)R_1(t)=\tilde b_2(r)R_1(t)R_2^\prime.
 \label{34bis}
\end{equation}

Then reparametrising $r$ as $\tilde b_2(r)dr\Rightarrow dr$, we may put without loss of generality $B=R_1(t)R_2(r)^\prime$, or equivalently $Z=1$. Thus, it appears  that  all non--dissipative  configurations evolving homologously and satisfying $Y_{TF}=0$, belong to what are called ``Euclidean stars'' \cite{euc}, characterized by the condition $Z=1 \Rightarrow B=R^\prime$. However as we shall see below, among all possible solutions satisfying the ``Euclidean condition'' , only one evolves homologously and satisfies the condition $Y_{TF}=0$.

Indeed,  we may rewrite  the field equations (\ref{13}), (\ref{14}) and (\ref{15}) as
\begin{equation}
4\pi q=-\frac{\dot Z}{Z^2R},
\label{necnd1}
\end{equation}
\begin{equation}
8\pi (P_r-P_{\bot})=\frac{\dot Z \dot R}{Z R}+\frac{1}{Z^2 R^2}\left(\frac{Z^\prime R}{ZR^\prime}+1-Z^2\right).
\label{necnd2}
\end{equation}

Since in this case we have $Z=1$ then $\Pi=P_r-P_\bot=0$ which implies because of the $Y_{TF}=0$ condition, that $\mu^\prime=0$. 

But we know that a shear--free, geodesic (non--dissipative) fluid with isotropic pressure is necessarily dust with homogeneous energy density and vanishing Weyl tensor (see \cite{gen, 20}). It goes without saying that this kind of system represents the simplest possible configuration (Friedman--Robertson--Walker).

Thus for the non--dissipative case, the homologous condition implies  $Y_{TF}=0$ and produces the simplest configuration. This configuration is the only one evolving homologously and satisfying  $Y_{TF}=0$. 

Of course, solutions satisfying $Y_{TF}=0$   but not evolving homologously do exist. They only require  $8\pi \Pi=\frac{4\pi}{R^3}\int^r_0{R^3 \mu^\prime dr}$. In such a case the solutions are shearing, and neither conformally flat nor--geodesic. 

Based on all the precedent comments we shall assume the homologous evolution as the simplest one. It is worth recalling that in the non--dissipative case both  conditions (homologous and homogeneous expansion) imply each  other.

\subsection{The dissipative case}
In the dissipative case, we may obtain from (\ref{5b1}) and (\ref{33bis}), 

\begin{equation}
\dot \sigma =-Y_{TF}+\left(\frac{ \dot R}{R}\right)^2-\left(\frac{ \dot B}{B}\right)^2.
\label{37bis}
\end{equation}

Then, taking the $t$-derivative of (\ref{con4}) and using (\ref{37bis}) we obtain
\begin{equation}
Y_{TF}\frac{R^\prime}{R}=4\pi Bq\left(\frac{\dot{q}}{q}+2\frac{\dot B}{B}+\frac{\dot R}{R}\right).
\label{38bis}
\end{equation}

If we assume $Y_{TF}=0$, then  we obtain
\begin{equation}
 q=\frac{f(r)}{B^2R},
\label{39bis}
\end{equation}
implying
\begin{equation}
 \dot q=-q(\Theta+\sigma),
\label{39bisb}
\end{equation}
where $f$ is an arbitrary integration function.
Solutions of this kind may be found by using the general methods presented in \cite{TM, TMb, ivanov1, ivanov}.

Now, in a dissipative process,  the stationary state, i.e. the absence of transient phenomena might be regarded as an  example of the simplest dissipative regime. Thus, if we assume the stationary state (neglecting the relaxation time), then the transport equation (\ref{tre}) reads
\begin{equation}
q=-\frac{\kappa T^\prime}{B}.
\label{40bis}
\end{equation}

Combining the above equation with (\ref{39bis}) we obtain
\begin{equation}
T^\prime=-\frac{f(r)}{\kappa BR}.
\label{41bis}
\end{equation}

At this point, however, neither can we  support further the assumption about the vanishing of the relaxation time as an indicator of  minimum complexity about the dissipative regime, nor can we prove that exact solutions of this kind exist.
\section{Stability of the vanishing complexity factor condition}
Using the general method developed in \cite{20} we may write  evolution equations for the structure scalars [see eq.(102) in that reference]. Thus,  with the notation used here, we obtain for the evolution of $X_{TF}$ the equation (\ref{g'}), which produces

\begin{eqnarray}
&-&4\pi( \mu+P_r)\sigma-\frac{4\pi}{B}\left( q^\prime-\frac{q R^\prime}{R}\right)\nonumber \\&-&\dot Y_{TF}-8\pi\dot \Pi-\frac{3\dot R Y_{TF}}{R}-16\pi \Pi\frac{\dot R}{R}=0,
\label{sta1}
\end{eqnarray}
where (\ref{EY}), (\ref{EX})  and (\ref{j4}) have been used.

Our goal in this section consists in looking for the conditions under which an initial state of vanishing complexity factor propagates in time under the homologous condition.

Let us first consider the non--dissipative case. In this latter case we have at some initial moment (say $t=0$) $Y_{TF}=q=\sigma=\Pi=0$,  then  (\ref{sta1}) becomes

\begin{eqnarray}
-\dot Y_{TF}-8\pi\dot \Pi=0.
\label{sta2}
\end{eqnarray}

It is worth noticing that taking the $t$-derivative of (\ref{Y}) and evaluating at $t=0$, under the conditions above it follows that $(\dot \mu)^\prime=0$.

Next, taking the $t$-derivative of (\ref{sta1}), and recalling that if $Y_{TF}=0$ the only solution compatible with an initially shear--free flow is a shear--free flow (see \cite{26} for details) we obtain
\begin{eqnarray}
-\ddot Y_{TF}-8\pi\ddot \Pi+\frac{8\pi \dot \Pi \dot R}{R}=0.
\label{sta3}
\end{eqnarray}

On the other hand, taking the second time derivative of  (\ref{Y}) and using (\ref{sta3}) it follows that
\begin{eqnarray}
2\dot R\dot \Pi=\frac{1}{R^2}\int^r_0(\ddot \mu)^\prime dr.
\label{sta3b}
\end{eqnarray}

Continuing with this process and taking the $t$-derivative of order $n$ (for any $n>0$) we see that the system could depart from the vanishing complexity factor condition only if it departs from the isotropic pressure  and the homogeneous energy density conditions. Deviations from these two conditions   are, of course, related as indicated in (\ref{sta3b}).

In the most general case, when the system is dissipative, we have at the initial moment 
\begin{eqnarray}
&-&4\pi( \mu+P_r)\sigma-\frac{4\pi}{B}\left( q^\prime-\frac{q R^\prime}{R}\right)\nonumber \\&-&\dot Y_{TF}-8\pi\dot \Pi-16\pi \Pi\frac{\dot R}{R}=0.
\label{sta4}
\end{eqnarray}

Without entering into a detailed discussion about this last equation, let us just mention the obvious fact that now the heat flux also affects the stability of the $Y_{TF}=0$ condition.
\section{conclusions}
We have discussed about the complexity of dynamical spherically symmetric relativistic fluid distributions. For doing that we have considered  two different (although related) aspects of  the definition of complexity when dealing with a dynamical fluid. On the one hand we have considered the problem of measuring the complexity of the structure of the fluid itself, and on the other we have  considered the degree of complexity of the pattern of evolution of the fluid distribution. 

As a measure of complexity of the structure of the fluid (the complexity factor) we have chosen the scalar function $Y_{TF}$. The reasons for doing so are the following:
\begin{enumerate}
\item It is the same complexity factor as for the static case, ensuring thereby that in the limit to the static regime we recover the correct expression for the complexity factor.
\item It encompasses the dissipative variables.
\item In the non--dissipative case, the homologous condition implies the vanishing of $Y_{TF}$.
\end{enumerate}

Next, we discussed about the complexity of the pattern of evolution. Two possibilities appear as the more obvious candidates: the homologous condition and the homogeneous expansion. We have leaned  toward the former option, for the following reasons:
\begin{enumerate}
\item It implies that the fluid is geodesic, even in the most general (dissipative) case. It is clear that the geodesic flow represents one of the simplest patterns of evolution.
\item In the non--dissipative case it implies that $Y_{TF}=0$, meaning that (in this case) the simplest pattern of evolution already implies the simplest structure of the fluid distribution.
\item In the non--dissipative case it leads to a unique solution, which from simple physical analysis appears as the simplest possible system.
\end{enumerate}

Next, we tackled the problem of the stability of the vanishing   complexity factor condition. In the non--dissipative case it appears clearly that such a condition will propagate in time, as far as the pressure remains isotropic. In the dissipative case, however, the situation is much more complicated and dissipative terms also may deviate the system from the vanishing complexity factor condition.

Finally we point out, what we believe is the main unsolved problem (in the spherically symmetric case). Indeed, in the dissipative case we have found that the heat flux vector satisfying the vanishing complexity factor condition is of the form given by the equation (\ref{39bis}). However in spite of all the efforts deployed so far, important questions remain unanswered, namely:
\begin{enumerate}
\item Do physically meaningful dissipative models satisfying  (\ref{39bis}) exist?
\item If the answer to the above question is positive then, is there a unique solution or  are there a large number of them?
\item What is the physical meaning of such solution(s)?
\item Is it physically reasonable to neglect transient effects when considering the simplest dissipative system, and assume that the relaxation time vanishes?
\item To summarize the questions above: is there a specific dissipative regime that could be considered as the simplest one?
\end{enumerate}

Besides the questions above, there is an obvious pending problem, regarding the extension of these results to non--spherically symmetric fluid distributions.

\begin{acknowledgments}

This  work  was partially supported by the Spanish  Ministerio de Ciencia e
Innovaci\'on under Research Projects No.  FIS2015-65140-P (MINECO/FEDER).

\end{acknowledgments}
\
\appendix
\section{The charged case}
If we assume the fluid to be electrically charged, then we have for $Y_{TF}$ (see \cite{21} for details)

\begin{equation}
Y_{TF}= - 8 \pi \Pi^{eff} + \frac{4 \pi}{R^3}\int^r_0{R^3\left( \mu_{eff}^{\prime} - \frac{3  q BU}{R}\right)dr},
\label{Yieff}
\end{equation}
with 
\begin{equation}
 \mu_{eff} =  \mu + \frac{s^2}{8\pi R^4},
\label{mueff}
\end{equation}
\begin{equation}
 P^{eff}_ r= P_r
 -\frac{s^2}{8\pi R^4},
\label{T11} 
\end{equation}
\begin{equation}
 P^{eff}_\bot
=P_{\perp}
 +\frac{s^2}{8\pi R^4},
 \label{T22}
\end{equation}
and
\begin{equation}
P^{eff}_r-P^{eff}_{\bot}\equiv \Pi ^{eff} = \Pi  - \frac{s^2}{4 \pi R^4},
\label{pieff}
\end{equation}
where $s(r)$ denotes the electric charge interior to radius $r$, and is given by 
\begin{equation}
s(r)=4\pi\int^r_0\varsigma BR^2dr, \label{13}
\end{equation}
where   $\varsigma$, is the charge density.

Thus, $Y_{TF}$ has the same form as for the neutral fluid, with  the physical variables $\mu, P_r, P_\bot$ replaced by their corresponding ``effective variables'' (\ref{mueff})--(\ref{pieff}). As a matter of fact, all the relevant equations are the same modulo this replacement (see \cite{sb} for a detailed treatment of this case in the static regime).
\section{Einstein equations}
 Einstein's field equations for the interior spacetime (\ref{1}) are given by
\begin{equation}
G_{\alpha\beta}=8\pi T_{\alpha\beta},
\label{2}
\end{equation}
and its non zero components
with (\ref{1}), (\ref{3}) and (\ref{5})
become
\begin{widetext}
\begin{eqnarray}
8\pi T_{00}=8\pi  \mu A^2
=\left(2\frac{\dot{B}}{B}+\frac{\dot{R}}{R}\right)\frac{\dot{R}}{R}
-\left(\frac{A}{B}\right)^2\left[2\frac{R^{\prime\prime}}{R}+\left(\frac{R^{\prime}}{R}\right)^2
-2\frac{B^{\prime}}{B}\frac{R^{\prime}}{R}-\left(\frac{B}{R}\right)^2\right],
\label{12} \\
8\pi T_{01}=-8\pi qAB
=-2\left(\frac{{\dot R}^{\prime}}{R}
-\frac{\dot B}{B}\frac{R^{\prime}}{R}-\frac{\dot
R}{R}\frac{A^{\prime}}{A}\right),
\label{13} \\
8\pi T_{11}=8\pi P_r B^2 
=-\left(\frac{B}{A}\right)^2\left[2\frac{\ddot{R}}{R}-\left(2\frac{\dot A}{A}-\frac{\dot{R}}{R}\right)
\frac{\dot R}{R}\right]
+\left(2\frac{A^{\prime}}{A}+\frac{R^{\prime}}{R}\right)\frac{R^{\prime}}{R}-\left(\frac{B}{R}\right)^2,
\label{14} \\
8\pi T_{22}=\frac{8\pi}{\sin^2\theta}T_{33}=8\pi P_{\perp}R^2
=-\left(\frac{R}{A}\right)^2\left[\frac{\ddot{B}}{B}+\frac{\ddot{R}}{R}
-\frac{\dot{A}}{A}\left(\frac{\dot{B}}{B}+\frac{\dot{R}}{R}\right)
+\frac{\dot{B}}{B}\frac{\dot{R}}{R}\right]\nonumber \\
+\left(\frac{R}{B}\right)^2\left[\frac{A^{\prime\prime}}{A}
+\frac{R^{\prime\prime}}{R}-\frac{A^{\prime}}{A}\frac{B^{\prime}}{B}
+\left(\frac{A^{\prime}}{A}-\frac{B^{\prime}}{B}\right)\frac{R^{\prime}}{R}\right].\label{15}
\end{eqnarray}
\end{widetext}
The component (\ref{13}) can be rewritten with (\ref{5c1}) and
(\ref{5b}) as
\begin{equation}
4\pi qB=\frac{1}{3}(\Theta-\sigma)^{\prime}
-\sigma\frac{R^{\prime}}{R}.\label{17a}
\end{equation}
\section{Dynamical equations}

The non trivial components of the Bianchi identities, $T^{\alpha\beta}_{;\beta}=0$, from (\ref{2}) yield
\begin{widetext}
\begin{eqnarray}
T^{\alpha\beta}_{;\beta}V_{\alpha}=-\frac{1}{A}\left[\dot { \mu}+
\left( \mu+ P_r\right)\frac{\dot B}{B}
+2\left( \mu+P_{\perp}\right)\frac{\dot R}{R}\right] 
-\frac{1}{B}\left[ q^{\prime}+2 q\frac{(AR)^{\prime}}{AR}\right]=0, \label{j4}\\
T^{\alpha\beta}_{;\beta}\chi_{\alpha}=\frac{1}{A}\left[\dot { q}
+2 q\left(\frac{\dot B}{B}+\frac{\dot R}{R}\right)\right] 
+\frac{1}{B}\left[ P_r^{\prime}
+\left(\mu+ P_r \right)\frac{A^{\prime}}{A}
+2( P_r-P_{\perp})\frac{R^{\prime}}{R}\right]=0, \label{j5}
\end{eqnarray}
\end{widetext}
or, by using (\ref{5c}), (\ref{5c1}), (\ref{16}), (\ref{23a}) and (\ref{20x}), they become, respectively,
\begin{widetext}
\begin{eqnarray}
D_T \mu+\frac{1}{3}\left(3 \mu+ P_r+2P_{\perp} \right)\Theta 
+\frac{2}{3}( P_r-P_{\perp})\sigma+ED_R q
+2 q\left(a+\frac{E}{R}\right)=0, \label{j6}\\
D_T q+\frac{2}{3} q(2\Theta+\sigma)
+ED_R  P_r 
+\left( \mu+ P_r \right)a+2(P_r-P_{\perp})\frac{E}{R}=0.
\label{j7}
\end{eqnarray}
\end{widetext}
This last equation may be further tranformed as follows, the acceleration $D_TU$ of an infalling particle can
be obtained by using (\ref{5c}), (\ref{14}), (\ref{17masa})  and (\ref{20x}),
producing
\begin{equation}
D_TU=-\frac{m}{R^2}-4\pi  P_r R
+Ea, \label{28}
\end{equation}
and then, substituting $a$ from (\ref{28}) into
(\ref{j7}), we obtain
\begin{widetext}
\begin{eqnarray}
\left( \mu+ P_r\right)D_TU 
=-\left(\mu+ P_r \right)
\left[\frac{m}{R^2}
+4\pi  P_r R\right] 
-E^2\left[D_R  P_r
+2(P_r-P_{\perp})\frac{1}{R}\right] 
-E\left[D_T q+2 q\left(2\frac{U}{R}+\sigma\right)\right].
\label{3m}
\end{eqnarray}
\end{widetext}
in terms of $Y_{TF}$ we may write (\ref{3m}) as
\begin{widetext}
\begin{eqnarray}
\left(\mu+ P_r \right)D_TU 
=-\left( \mu+P_r \right)4\pi R
\left[\frac{ \mu}{3}
+ P_r-\frac{1}{3R^3}\int^r_0{R^3\left(  \mu^{\prime} - \frac{3  q BU}{R}\right)dr}\right] \nonumber\\
-E^2\left[D_R  P_r
-\frac{Y_{TF}}{4\pi R}+\frac{1}{R^4}\int^r_0{R^3\left( \mu^{\prime} - \frac{3  q BU}{R}\right)dr}\right] 
-E\left[D_T q+2 q\left(2\frac{U}{R}+\sigma\right)\right].
\label{3mb}
\end{eqnarray}
\end{widetext}
\section{Evolution of structure scalars}
From the Bianchi identities evolution equations for the structure scalars can be derived (see (102) in \cite{20}, or (58) in \cite{H5}).  Specifically, for $X_{TF}$ one obtains
\begin{eqnarray} 
(4\pi  \mu+X_{TF}\dot
)&+&\frac{1}{3}(2X_{TF}+Y_T+X_T-Y_{TF})(\Theta-\sigma)A\nonumber \\&+&12\pi
q\frac{AR^\prime}{BR}=0.\label{g'}
\end{eqnarray}
\section{The transport equation}
Assuming a causal dissipative theory (e.g.the Israel-- Stewart theory \cite{20n, 21n} ) the transport equation for the heat flux reads
\begin{eqnarray}
\tau h^{\alpha \beta}V^\gamma q_{\beta;\gamma}&+&q^\alpha=-\kappa h^{\alpha \beta}\left(T_{,\beta}+Ta_\beta\right)\nonumber \\&-&\frac{1}{2}\kappa T^2 \left(\frac{\tau V^\beta}{\kappa T^2}\right)_{;\beta} q^\alpha.
\label{tre}
\end{eqnarray}
where $\kappa$ denotes the thermal conductivity, and $T$ and $\tau$ denote temperature and relaxation time respectively.

\end{document}